\documentclass[twocolumn,showpacs,amsfonts,aps,prl,superscriptaddress]{revtex4}

\usepackage{graphics,epsfig}
\usepackage{bm}
\usepackage{dsfont}
\usepackage{amssymb}

\begin{document}

\newcommand{\XXX}[1]{}

\newcommand{\grad}{\bm{\nabla}}
\newcommand{\Grad}{\bm{\nabla}_{\!3}}
\newcommand{\vct}[1]{\mathbf{#1}}
\newcommand{\uvct}[1]{\mathbf{\hat{#1}}}
\newcommand{\tens}[1]{\mathbf{#1}}
\newcommand{\kruemm}{{\kappa}}
\newcommand{\Nabla}[1]{\nabla_{\!\! {#1}}}
\newcommand{\Order}[2]{\mathcal{O}\left({#1}^{#2}\right)}
\newcommand{\deh}{\delta\!h}
\newcommand{\erf}{\text{erf}}

\def\R{\rm I\kern-.18em R}
\def\be#1{\begin{equation}  \label{#1}}
\def\ee{\end{equation}}
\def\nor#1{\vert #1 \vert}
\def\intdrr{\int\!\!\!\!\int\limits_{\!\!\!\Omega}\!\!}
\def\mm{\;{\rm mm}}
\def\mum{\mu{\rm m}}
\def\text#1{{\rm #1}}

\title{Thermal noise influences fluid flow in thin films
during spinodal dewetting}

\author{R. Fetzer}
\affiliation{Department of Experimental Physics, Saarland
University, 66041 Saarbr\"ucken, Germany}
\altaffiliation[Present address:]{Ian Wark Research Institute,
University of South Australia, Mawson Lakes, SA 5095, Australia}
\author{M. Rauscher}
\affiliation{Max-Planck-Institut f\"{u}r Metallforschung,
Heisenbergstr.\ 3, 70569 Stuttgart, Germany}
\author{R. Seemann}
\affiliation{Max-Planck-Institut f\"{u}r  Dynamik und
Selbstorganisation, Bunsenstr. 10, 37073 G\"ottingen, Germany}
\author{K. Jacobs}
\affiliation{Department of Experimental Physics, Saarland
University, 66041 Saarbr\"ucken, Germany}
\author{K. Mecke}
\affiliation{ Institut f\"ur Theoretische Physik,  Universit\"at
Erlangen-N\"urnberg, Staudtstrasse 7, 91058 Erlangen,  Germany }
\email{mecke@physik.uni-erlangen.de}

\date{\today}


\begin{abstract}
Experiments on dewetting thin polymer films confirm the
theoretical prediction that thermal noise can strongly influence
characteristic time-scales of fluid flow and cause coarsening of
typical length scales. Comparing the experiments with
deterministic simulations, we show that the Navier-Stokes equation 
has to be extended by a conserved bulk noise term to accomplish the 
observed  spectrum of capillary waves. Due to thermal fluctuations the spectrum 
changes  from an exponential 
to a power law decay  for large wavevectors.  Also
the time evolution of the typical wavevector of unstable
perturbations exhibits noise induced coarsening that is absent in
deterministic hydrodynamic flow.
\end{abstract}

\maketitle
With the advent of nanofluidics in the last 
 years  it became evident that thermal noise may play an important role  
in all hydrodynamic processes occuring at free interfaces on 
small scales. Although in bulk fluids  
hydrodynamic Navier-Stokes 
equations are proven to be valid down to the nanometer scale, in free interface flow 
stochastic forces induced by  molecular motion can significantly alter the behavior 
even on a micrometer scale.  
Moseler and Landman, for instance,  found that the  
deterministic lubrication approximation for axial-symmetric free 
boundary flow is not applicable 
for the description of nanoscopic cylindrical jets \cite{moseler00}. 
The lack of thermally triggered fluctuations 
in the classical hydrodynamic continuum modelling was identified as the most 
likely source for deviations of Navier-Stokes equation from 
molecular dynamics simulations.  
They derived a stochastic differential equation that includes 
thermal noise, whose influence on the dynamics increases as 
the radius of the nanojet becomes smaller, leading finally to 
the emergence of symmetric double cone neck shapes during the breakup 
instead of a long thread solution as expected in the absence of noise.  
In Ref. \cite{eggers02}  path integral methods were applied to confirm 
that thermal noise induces indeed qualitative changes in the breakup of 
a liquid nanometer jet: Thermal fluctuations 
speed up 
the dynamics and make surface tension an irrelevant force for 
the breakup. Very recently, the importance of thermal noise for drop formation 
 was observed in a colloidal dispersion with an ultra-low 
surface tension \cite{hennequin06}.

Dewetting of thin films is another technologically
important free surface flow problem 
where a  lubrication approximation is used widely. 
So far, thin film flow has
been studied by hydrodynamic Navier-Stokes equations without
considering thermal noise, e.g., \cite{oron97}. With
decreasing film thickness, however, it is expected that thermal
noise gains significance \cite{mecke05}. In
Refs.~\cite{mecke05,gruen05,davidovitch05}, thermal fluctuations
have been taken into account and a stochastic version of the thin
film equation was derived based on the lubrication approximation
for stochastic hydrodynamic equations \cite{landau91b,fox70a}.   
The magnitude of the added stochastic forces increases as the thickness 
of the film decreases. 
Recent analytical and 
numerical studies of the stochastic thin film evolution indicate  that
thermal noise  increases  characteristic timescales of the
dewetting process \cite{gruen05} and changes  qualitatively the time evolution of the
film thickness $h(\vct{r},t)$ (with $\vct{r}=(x,y)$),
i.e., its power spectrum $\tilde{C}(q,t)$ \cite{mecke05}. 
Here, we report 
that these predictions of noisy hydrodynamics can be
confirmed experimentally by in situ atomic force microscopy
(AFM) of dewetting thin polymer films.

\begin{figure}
\centerline{ \epsfig{file=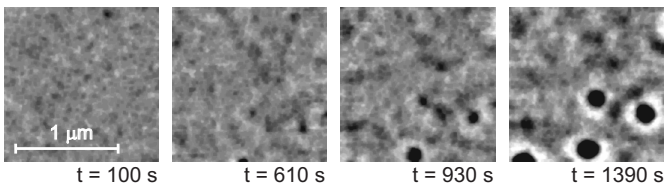, width=8.8cm}}
\caption{Dewetting of a 3.9~nm polystyrene film as monitored by in situ
AFM at 53$^\circ$C\/. \label{movie} }
\end{figure}

For the experiments, we used polystyrene (PS) of 2.05~kg/mol molecular
weight,
prepared on a silicon wafer with a 191~nm thick
silicon oxide layer. By spin casting a toluene solution of PS we
prepared films with thicknesses of $(3.9\pm 0.5)$~nm and $(4.6\pm 0.5)$~nm 
(as measured by ellipsometry at different sites). These
films are unstable and dewet spinodally. Above the glass transition
temperature, the PS film is liquid and capillary waves can be
observed \cite{seemann01}\/. 
Below a characteristic wave number their amplitudes grow
exponentially in time, eventually leading to holes in the film
\cite{Ruckenstein,Herminghaus}\/. Dewetting then
proceeds by the growth of holes and their coalescence. 
By tapping mode$^{\rm TM}$ AFM, the entire dewetting
process can be monitored in situ, c.f. Fig.~\ref{movie}.

In addition to the experiments, we use deterministic simulation
data of the spinodal dewetting process from Ref.~\cite{becker03}\/.
In these simulations, the experimental system parameters film thickness,
viscosity, and effective interface potential $\Phi(h)$ 
were used. 
We gain quantitative information about the influence of thermal
noise to the dewetting process by analyzing the variance
$\sigma^2(t)=\overline{h^2}-\overline{h}^2$ of the film height
$h(\vct{r},t)$ and the variance $k^2(t)= \overline{(\vct{\nabla}
h)^2}/ (2\pi\sigma^2(t))$ of the local slope of $h(\vct{r},t)$.
The quantity $k^2(t)$ gives information about the 
preferred wavevector within the film surface in the early
stage of dewetting before holes are formed, cf.
Fig.~\ref{movie}\/.  
In this early stage regime, the overbars represent an integration
over all positions of the image. During the later
stage, however, when holes appear and grow, the linearization of
the thin film equation (TFEq) \cite{mecke05},
which we use to analyse the data, is in general not valid, but may
still be used in between the holes, where the film
heights are close to the prepared film thickness $h_0$\/. We
therefore analyse in the later stage of dewetting the film only at
regions where $h(\vct{r},t)\approx h_0$ and ignore the other parts.
Note that this spatially selective data analysis is not possible by
using correlation function or Fourier space techniques.

A main problem in the data analysis is the finite size of 
the observation window and the emergence of growing holes in
the thin film which limits the film regions where linear dynamics 
can be studied.    
The variances $\sigma^2$ and $k^2$ (which are moments of the
spectrum) cannot be determined accurately from the Fourier
transform of the AFM images.
The measured spectrum is convoluted with the Fourier transform of
the observation window, which decays 
as $\sim q^{-2}$ for a sharp cutoff.
Thus, the short-wavelength behavior of the spectrum is in our case
dominated by the edges of the AFM image and not by the thermal
fluctuations.
Moreover, the exponentially damped spectrum $\tilde{C}\sim
e^{-2t|\omega(q)|}$, characteristic for the deterministic dynamics,
is masked by this convolution yielding a spectrum similar to the
algebraic noise-induced 
$\tilde{C}_{CW}(q) \sim q^{-2}$ \cite{mecke05}. 
Since the 'bare' spectrum $\tilde{C}(q)$ is 
experimentally not accessible,
one has to determine variances $\sigma^2$ and $k^2$ 
from the real space images.
Whereas determining $\sigma^2$ is straightforward, 
calculating $k^2$ from pixelized images using numerical
differentiation is too inaccurate.

\begin{figure}
\includegraphics[width=0.48\linewidth]{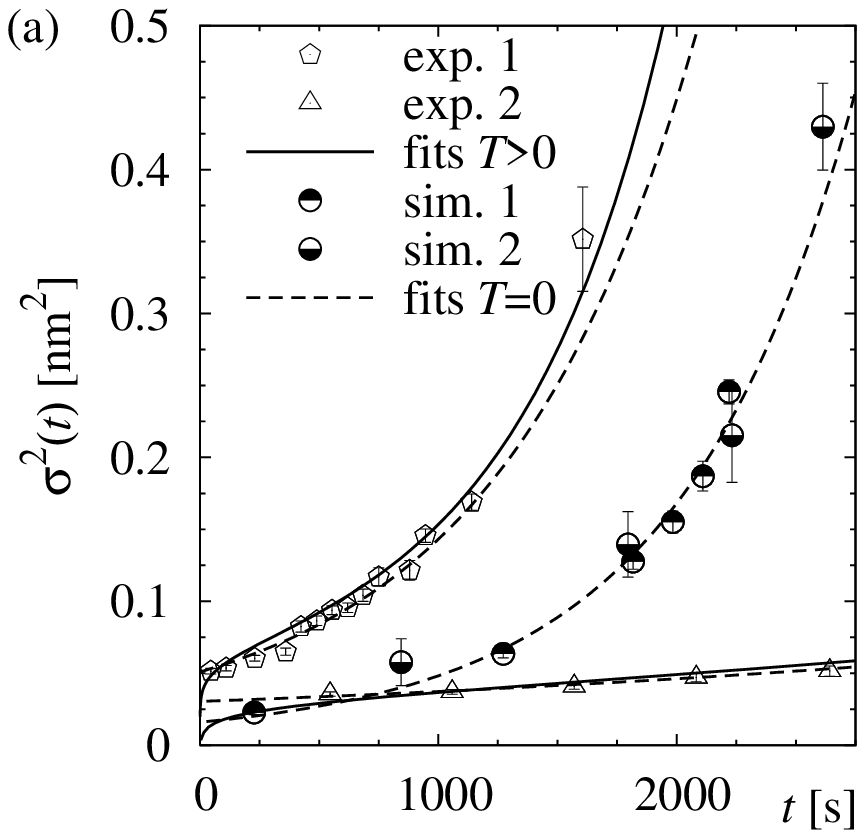}
\includegraphics[width=0.48\linewidth]{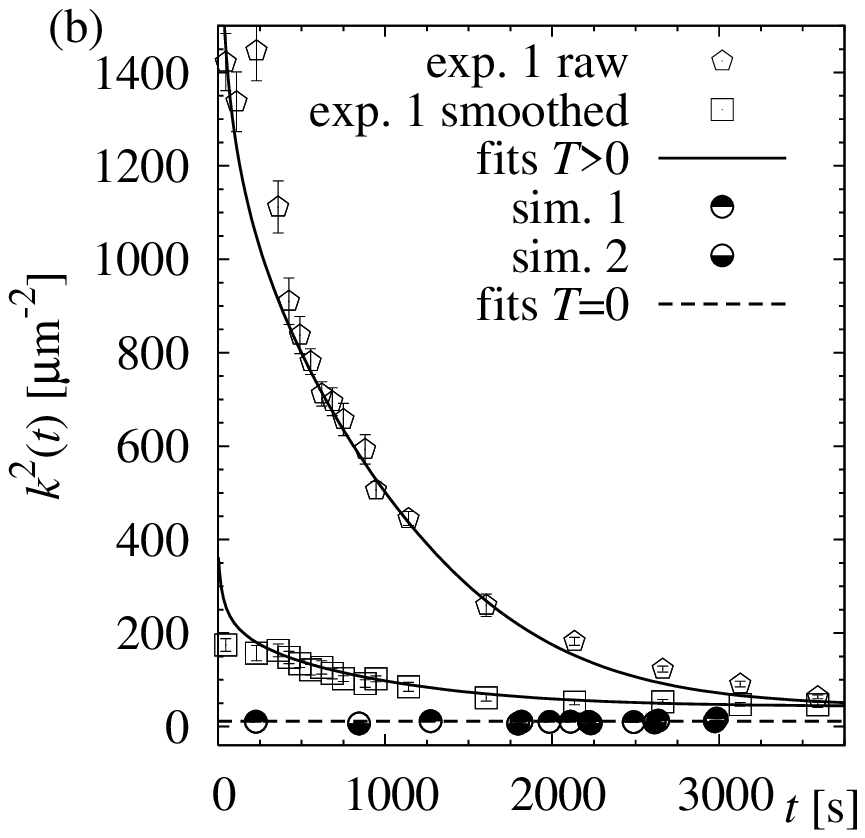}
\caption{Analysis of AFM experiments and of deterministic 
simulations ($T=0$):
(a) The roughness $\sigma^2(t)$ and (b) the
variance $k^2(t)$ of the local slope (only exp.~1 and simulations for clarity) 
as functions of time and fits
based on the linearized TFEq.
While 
$\sigma^2(t)$ can be fitted with the deterministic theory by
adjusting the initial roughness $\sigma_0^2$ and the
characteristic time scale $t_0$, this is not
possible for the experimental $k^2(t)$\/. In the linear regime the deterministic $k^2(t)$ is constant in time. 
\label{fig-time}}
\end{figure}

Here, we apply an analysis technique based on Minkowski functionals
$M_\nu$ \cite{becker03}. 
We threshold the images at a certain film height $h$ 
and measure the area $M_0(h)$,  boundary length 
$M_1(h)$, and Euler number $M_2(h)$ as function of $h$.  
In the linear regime of spinodal dewetting,
the fluctuations in these quantities should follow a Gaussian 
distribution which was previously tested in Ref.~\cite{becker03}. 
Then, the mean values of $M_\nu(h)$ depend only on the variances
$\sigma^2$ and $k^2$, which can be determined 
as fit parameters to the functions  $M_\nu(h)$. 
In Fig.~\ref{fig-time} we show the resulting temporal evolution of
$\sigma^2(t)$ and of $k^2(t)$ in the linear regime. 
Besides providing accurate estimations of these values, 
this morphometric analysis allows to test whether the analysed film region 
is still in the linear regime. 
As soon as the non-linear dynamics 
sets in, 
the Gaussian model 
fails to fit 
the measured functions $M_\nu(h)$. 
We compare the data 
from  two experiments 
at two spatial resolution lengths  
(exp.~1: pixel size $r_p\approx 6\:\text{nm}$ temperature $T\approx 53^o$C;
exp.~2: $r_p\approx 40\:\text{nm}$, 
$T\approx 65^o$C)  
\XXX{SCANGEROESSE IST HIER IRRELEVANT, $r_m\approx r_p$, HAT AUF
SEITE 3 NOCH GEFEHLT.}
with data from 
deterministic simulations at similar physical conditions and find substantial
differences. 
In the following, we 
explain
these differences by
introducing thermal noise to the description of thin film flow.

In both experiments the short chain length PS film can be described as an
incompressible Newtonian liquid (with a 
shear viscosity on the order of $\eta\approx 10^4$~Ns/m$^2$) on an
infinite, flat solid substrate. Thus, the flow can be modelled by a stochastic
Navier-Stokes equation with an additional  random stress fluctuation tensor 
representing the effect of thermal
molecular motion \cite{landau91b,fox70a}. 
We assume 
no-slip boundary conditions between liquid
and substrate and full-slip at the liquid/air interface
$z=h(\vct{r},t)$\/. At the latter, the normal stress is balanced
by the surface tension $\gamma$ and the disjoining pressure
$-\Phi'$\/.
For a smooth thin film, where  the characteristic film height
$h_0\approx 4$~nm is much smaller than the length scale $\approx
100$~nm  over which the film thickness varies laterally, 
a long wavelength expansion yields the following stochastic TFEq
\cite{oron97,mecke05,gruen05}
\begin{equation}
\label{finalequation} 
\frac{\partial h}{\partial t} = \grad
\left(\frac{h^3}{3\,\eta} \grad \left[\Phi'(h)-\gamma\Delta 
h\right] + \sqrt{\frac{2k_BT h^3}{3\,\eta}}\,\vct{{N}}(t)
\right)
\end{equation}
with a single multiplicative
conserved noise vector $\vct{{N}}(\vct{r},t)$ obeying
$\left\langle \vct{{N}}(\vct{r},t) \right\rangle =0$ and  
$\left\langle {N}_i(\vct{r},t) \,{ N}_j(\vct{r}',t')
\right\rangle = \delta_{ij}\,\delta(\vct{r}-\vct{r'})\,
\delta(t-t')$, where $\langle \cdot\rangle$ denotes
the ensemble average over 
realizations of the noise. 
The characteristic lateral length
scale for a thin film of $h_0 =4$~nm  is given by the spinodal wavelength $2\pi/q_0 =
\sqrt{-8\pi^2\gamma/\Phi''(h_0)} =
4\,h_0^2\,\sqrt{\pi^3\,\gamma/A}$, where the Hamaker constant $A$
determines the effective interface potential $\Phi(h) =
-\frac{A}{12\,\pi\,h^2}$. We
have ${2\pi \over q_0}\approx 400$~nm\/ with the values $A\approx
2\cdot 10^{-20}$~Nm and $\gamma \approx 3\cdot 10^{-2}$~N/m
determined experimentally for the same system 
\cite{seemann01}. 
The experimental parameters lead to dimensionless
amplitudes of the noise
$\frac{3\,k_B\,T}{8\,\pi^2\,h_0^2\,\gamma}\approx 4\cdot 10^{-4}$
and $2\cdot 10^{-4}$ for exps.~1 and 2, respectively. 
 In the following we show that even this small current 
induced by thermal fluctuations 
can substantially influence the dewetting process.

Comparing the experiments with the numerical solution of
Eq.~(\ref{finalequation}) with $T=0$, we indeed find excellent
agreement of the spatial structures \cite{becker03}\/.
Nevertheless, the time scales do not match. 
This can be
illustrated by comparing the time dependence of $\sigma^2(t)$ and
$k^2(t)$ for the AFM data and the deterministic thin film
simulations,
see Fig.~\ref{fig-time}\/. The most significant
deviation is observed for $k^2(t)$ which varies in time for the
experiments, but stays constant for the deterministic simulations.
This mismatch in $k^2(t)$ is most prominent for short times, i.e.,
for small $\sigma^2(t)$\/. 
 In the following, we
quantitatively analyse the early stages of dewetting in a linear
approximation of Eq.~(\ref{finalequation}) to demonstrate the
relevance of thermal noise in hydrodynamics on small length
scales.

In the beginning of the dewetting process the deviations
$\deh(\vct{r},t) =h(\vct{r},t)-h_0$ from the initial film height
$h_0$ are small as compared to $h_0$\/. By expanding
Eq.~(\ref{finalequation}) in first order of $\deh$ and $\vct{{N}}$
we obtain a linear stochastic equation in Fourier space. Its
dispersion relation $\omega(q) =[1-(q^2/q_0^2-1)^2]/t_0$ 
has its maximum at $q_0^2=
-{\Phi''(h_0)}/({2\gamma})$ and a characteristic time scale
$t_0={3\eta}/({\gamma\,h_0^3\,q_0^4})$ (about 300~s and 1700~s for
exp.~1 and 2, respectively). The spectrum reads 
$\tilde{C}(q,t) = \langle\tilde{\deh}(\vct{q},t)\, \tilde{\deh}^*(\vct{q},t)\rangle 
= \tilde{C}_0(q)\,e^{2\omega(q)\,t}
+ \frac{k_BT\,h_0^3}{3\eta}\, \frac{q^2}{\omega(q)} \left[
e^{2\omega(q)\,t}  -  1 \right]$ with the initial  spectrum
$\tilde{C}_0(q)$ at $t=0$\/. 
We assume the initial spectrum to be constant 
$\tilde{C}_0(q)=\frac{2\pi}{q_0^2}\sigma_0^2$ for
$q<\sqrt{2}q_0$ and zero otherwise. 
The spectrum has necessarily  a
microscopic cutoff $q_{m}=2\pi/r_m\gg q_0$ at the scale $r_m$,
which is certainly larger than the fluid molecules.  The cutoff for
the measured spectrum is set by the experimental resolution,
e.g., by the pixel size $r_p$ or a smoothing length if the data is
post-processed. 
Due to a rapid buildup of noise  induced short wavelength
roughness on a microscopic time scale $t_m =
({q_0}/{q_{m}})^4\,t_0$, the short wavelength part of the initial
spectrum $\tilde{C}_0(q>q_0)$ is irrelevant for the film evolution on
the characteristic time scale $t_0\gg t_m$\/.

In the case of a
spinodally unstable film with $\Phi''(h_0)<0$ the dispersion
relation $\omega(q)$ is negative for $q>\sqrt{2}\,q_0$\/. This
leads to two important differences between deterministic and
stochastic dewetting. First, and largely independent of the
initial roughness, for $t\rightarrow \infty$ and $q>\sqrt{2}\,q_0$
one finds an exponentially decaying power spectrum $\tilde{C}(q,t)
\rightarrow \tilde{C}_0(q)\,e^{-2\,|\omega(q)|\,t}$ in the
deterministic dynamics ($T=0$) but we recover the algebraic
capillary wave spectrum $\tilde{C}(q,t) \rightarrow
\frac{k_BT\,h_0^3}{3\eta}\,\frac{q^2}{|\omega(q)|}\sim
\frac{k_BT}{\gamma\,q^2} = \tilde{C}_{CW}(q)$ for any finite
temperature $T$\/. And second, for sufficiently smooth initial
spectra which decay rapidly for $q > \sqrt{2} q_0$, the maximum of
the deterministic spectrum stays at $q_0$ for all times, while the
maximum of the stochastic spectrum approaches $q_0$ from above as
$t\to \infty$, see Ref.~\cite{mecke05} for details. This noise
generated coarsening process can last until non-linearities become
important, effectively masking the most typical feature of the
linear deterministic regime, namely that the maximum of the power
spectrum stays at the fixed wave number $q_0$.

In order to further illustrate  the spatio-temporal features of the
dynamics we calculate the  time evolution of the roughness
of the film  $\sigma^2(t)= \int \frac{d^2q}{(2\,\pi)^2}\,
\tilde{C}(q,t)$ shown in 
Fig.~\ref{fig-time}(a) for the experiments discussed above and the
corresponding deterministic simulation. Both, the experimental and
the deterministic data can be fitted with the deterministic
$\sigma_{T=0}^2(t)$ with parameters $t_0$ and $\sigma_0$ in
reasonable agreement with the experimental parameters. The
initial roughness $\sigma_0^2$ is experimentally  not accessible.
Thus, the time evolution of the roughness $\sigma^2(t)$ gives no
clear indication for the importance of thermal fluctuations for
the dewetting process.

\begin{figure}
\includegraphics[width=0.9\linewidth]{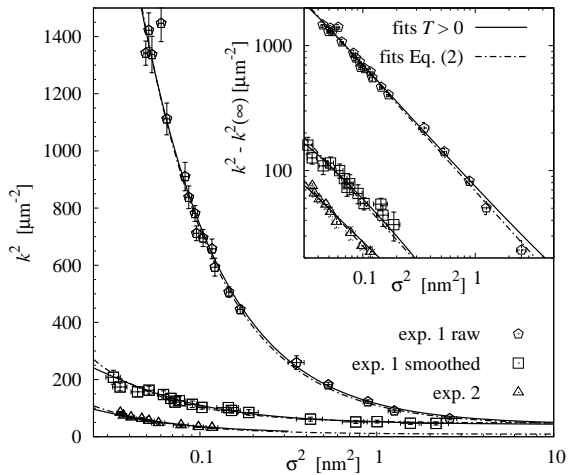}
\caption{
\label{fig-variances} 
Slope variance $k^2$  from the experiments plotted versus the roughness
$\sigma^2$\/. 
The solid lines are fits based on the lineraized stochastic TFEq,
whereas the  dash-dotted lines are asymptotic fits according to Eq.~(\protect\ref{coarsening}). 
The influence of noise is reduced for  smoothed as well as for
less resolved data (exp.2). 
The inset shows the same data in a double logarithmic plot
highlighting the validity of the asymptotic
Eq.~(\protect\ref{coarsening}).
 } 
\end{figure}

Yet, the variance of the local slope $2\pi\sigma^2(t)k^2(t) =
\langle\left[\grad \deh(\vct{r},t)\right]^2\rangle =
\int\frac{dq\,q^3}{2\,\pi}\tilde{C}(q,t)$ 
is
sensitive to thermal noise. We find 
a strong dependence on the noise term for times $t\sim t_0$ whereas the 
late stage $k^2(t\rightarrow\infty)=k_0^2$ is given by the maximum of 
the dispersion relation $k_0^2=
{q_0^2}/({2\,\pi})$ alone.  The reason why the noise term  
becomes irrelevant
for large $t$ is the strong
increase of $\sigma^2(t)$ with time due to the deterministic growth of unstable 
modes. Note that for a purely 
deterministic dynamics the variance $k^2_{T=0}(t) = k_0^2$ changes in time only at microscopic time 
scales $t_m$ and  remains constant in time if 
one chooses $q_m=\sqrt{2}q_0$.  
In particular, the position $q\approx k(t)$ of the
maximum in the structure function $\tilde{C}(q,t)$ does not change
during a deterministic dewetting process. In contrast, 
thermal noise induces a time dependence of $k^2(t)$\/.  
The experimental data cannot be fitted with the deterministic theory,
because $k^2(t)$ is clearly not constant in time. 
A simultaneous fit to the experimental data 
with $\sigma_0^2 =0$  yields
$k_BT/(4\pi\gamma)$
and the cut-off wavevector 
$q_m^2/q_0^2$
as given in 
Table~\ref{table}.
From the experimental parameters we would expect
$k_BT/(4\pi\gamma)\approx0.012\,\text{nm}^2$ for both experiments,
which is reasonably close to the fitted values.
The values for $q_m$ correspond to   
cutoff lengths  $r_m$ 
which are  larger than the nominal pixel size. 
The reason is probably that the experimental cutoff is smooth
rather than sharp.
However, the ratios of the $r_m$ for the different experiments 
matches quite well the ratios of the corresponding pixel sizes. 
In contrast to $\sigma_0^2$, smoothing changes $k^2$ drastically.
Deterministic simulation data is hardly affected by smoothing.
Smoothing obviously removes fluctuations on small scales, which are exponentially fast damped 
by the deterministic dynamics but are permanently (re)generated by
thermal noise in the experiments. 
A quantitative analysis of the experiments 
at different lateral resolution clearly  confirms thermal noise 
as the most likely source for the observed $k^2$ dependence. The 
dependence of $k^2$ on the upper cut-off  $q_m$ is a crucial test of  
the theoretical prediction as presented below in Eq. (\ref{coarsening}).

As the microscopic cutoff $q_{m}$ is much larger than $q_0$
there is a time regime $t>t_m\approx 10^{-2}t_0$
up to $t\approx t_0$ where $k^2(t)$ is approximately given by 
\begin{equation}
\label{coarsening}
k^2(t) \approx  
k_0^2 + \frac{\chi}{\sigma^2(t)},\quad \text{with}\quad
\chi = \frac{k_B\,T\,q_m^2}{8\,\pi^2\,\gamma},
\end{equation}
independent of the initial conditions \cite{mecke05}. Since $\sigma^2(t)$ is
monotonically increasing with time, the characteristic wavelength
$\sqrt{2\pi/k^2(t)}$ increases. Thus,
thermal noise generates coarsening even in the linear regime for
which the deterministic linear dynamics predicts a fixed
characteristic wavevector. 
Fig.~\ref{fig-variances} shows $k^2$ as a function of $\sigma^2$ 
fitted by Eq.~(\ref{coarsening}) as well as by the full linear
stochastic theory.
Both fits give consistent values for $k_0^2$ and $\chi$ which
are given in Table~\ref{table}\/. These values are also consistent
with the values determined by the system parameters given above.


From the fit values $k_0^2$ and the theoretical relation $q_0^2=
-{\Phi''(h_0)}/({2\gamma})$  we get for the film thicknes $h_0$ in exp.~1 
3.6~nm and 3.8~nm from the smoothed and raw data,
respectively, and for exp.~2 $h_0=5.5$~nm. These values are
in good agreement with our experimental data.
Accordingly, one finds from the fit values for $\chi$ the   cut-off 
lengths $r_m$ given in Table~\ref{table} --  in good agreement with the ratios of 
 pixels sizes and smoothing length, respectively. 
It is remarkable, that one can determine reliable 
values for film thicknesses and lateral resolution by measuring the effect of thermal 
noise on fluid flow.  
 This is a strong indication 
that the coarsening visible in $k^2(t)$ during the linear regime is indeed due to 
thermal fluctuations which can be described by the stochastic TFEq (\ref{finalequation}).    
Numerical solutions of Eq.~(\ref{finalequation}) would provide an excellent further test. 

In conclusion, the interplay of substrate potentials and thermal
noise 
can strongly influence characteristic time scales of fluid flow and leads to a
coarsening of typical lengths on a $\mu$m scale. 
The results will have impact
on the future development of 
nanofluidics,
enabling
reliable predictions of fluid flow in confined geometries such as
in liquid coatings or in lab-on-chip devices.

\begin{table}[h]
\begin{tabular}{l|c|c|c|c|c|c} 
dataset  &  $\frac{k_B T}{4\pi\gamma}$ [nm$^2$] & ${q_m^2\over q_0^2}$ & $k_0^2$ [$\mu$m$^{-2}$] & 
$\chi$
& $r_m$ [nm] \cr 
\hline 
1 smooth &  0.01 & 12 &  $48$ & $5.6\,10^{-6}$ & 116 \cr
1 raw & 0.01 & 170 & $40$  & $7\,10^{-5}$  & 33  \cr
2 raw & 0.008 & 28 & $9$  & $2.5\,10^{-6}$ & 174 
\end{tabular} 
\caption{\label{table} Experimental fit parameters for
Figs.~\protect\ref{fig-time} and \protect\ref{fig-variances}.}
\end{table}

It is a great pleasure to thank J. Becker and G. Gr{\"u}n
for granting  
access to the digital raw data of the simulations. 
Financial support by the DFG 
under grants Ja905/3, Se1118/2, Ra1061/2 and Me1361/9 
is gratefully acknowledged.

\end{document}